\documentclass[sigconf, nonacm]{acmart}
\AtBeginDocument{%
  }

\setcopyright{acmlicensed}
\copyrightyear{2026}
\acmYear{2026}
\acmDOI{XXXXXXX.XXXXXXX}
\acmConference[ICCAD '26]{International
Conference on Computer-Aided Design}{Nov 08--12, 2026}{San Jose, CA}
\acmISBN{978-1-4503-XXXX-X/2018/06}

\usepackage{algorithm}
\usepackage{algpseudocode} 
\usepackage{pifont}
\usepackage{amssymb} 
\usepackage{pifont}  
\usepackage{tabularx} 
\usepackage{booktabs} 

\newcommand{\xmark}{\textsf{x}} 
\usepackage{graphicx}
\usepackage{textcomp}
\usepackage[dvipsnames]{xcolor}
\usepackage{xspace}
\usepackage{subcaption}
\usepackage{tikz}
\usepackage{tabularx}
\usepackage{framed}
\usepackage{booktabs}
\usepackage{multirow}
\usepackage{xurl}
\usepackage{hyperref}
\usepackage{threeparttable}
\usepackage{stfloats}
\usepackage{placeins}
\usepackage{circledsteps}
\newif\ifdraft
\draftfalse

\newcommand{\Prv}{$\mathcal{P}$\xspace}
\newcommand{\Vrf}{$\mathcal{V}$\xspace}
\newcommand{\Cir}{$\mathcal{C}$\xspace}
\newcommand{\mPrv}{$\mathcal{P}^*$\xspace}
\newcommand{\sys}{\texttt{FASTAR}\xspace}
\begin{document}

\title{\sys: \underline{F}RI \underline{A}ccelerator for \underline{S}calable \underline{T}ransparent \underline{AR}guments of Knowledge\\}
\settopmatter{authorsperrow=2}



\author{Tengkai Gong}
\affiliation{%
  \institution{Northeastern University}
  \city{Boston}
  \country{USA}}
\email{gong.ten@northeastern.edu}

\author{Xiaolin Xu}
\affiliation{%
  \institution{Northeastern University}
  \city{Boston}
  \country{USA}
}
\email{x.xu@northeastern.edu}

\begin{abstract}
Zero-Knowledge Proofs (ZKPs) enable a prover to cryptographically convince a verifier of the validity of a statement without revealing any underlying secrets, forming a foundational primitive for verifiable computation. The ZKP landscape is undergoing a fundamental shift from classic zk-SNARKs such as Groth16, which rely on trusted setup and are vulnerable to quantum adversaries, toward transparent, post-quantum constructions such as zk-STARK. These systems achieve post-quantum security by relying solely on collision-resistant hash functions, however, at the cost of substantial computational overhead. In particular, the Fast Reed--Solomon Interactive Oracle Proof of Proximity (FRI) protocol dominates prover complexity, generating massive data volumes, repeated Merkle-tree commitments, and irregular memory access patterns that limit performance and energy efficiency on general-purpose processors.

To address these challenges, this work proposes \sys, a novel FPGA-based accelerator for the FRI protocol. Unlike accelerators that pursue fixed high-performance kernels on expensive ASIC process nodes, \sys adopts a constraint-driven design methodology. Our framework is implemented with High-Level Synthesis (HLS) and composed of fully parameterizable building blocks for the major stages of FRI, including polynomial evaluation, recursive split-and-fold, and Merkle-tree construction. From user-provided board specifications, \sys automatically generates hardware implementations tailored to the resource and memory constraints of the target FPGA, enabling deployment across a wide range of platforms without manual redesign. 

We evaluate \sys across three representative FPGA platforms: a high-end datacenter FPGA with High Bandwidth Memory (HBM), a mainstream embedded system-on-chip, and a low-power edge device. Experimental results show that \sys achieves up to $25.7\times$ and $3.5\times$ speedups over highly optimized CPU and GPU baselines, respectively, while delivering superior energy efficiency. These results demonstrate that, as a single modular hardware framework, \sys can scale across FPGA classes and make practical FRI acceleration feasible for real-world zk-STARK systems.
\end{abstract}

\ccsdesc[500]{Security and privacy~Cryptography}
\ccsdesc[500]{Computer systems organization~Architectures}
\keywords{Zero-Knowledge Proofs, Hardware Acceleration}

\received{14 April 2026}
\received[accepted]{11 July 2026}

\maketitle
\section{Introduction}

Zero-Knowledge Proofs (ZKPs) enable a prover \Prv to convince a verifier \Vrf of the correctness of a computation without revealing private inputs \cite{zkp}. This capability is valuable in distributed systems such as verifiable and privacy-preserving computation, where a resource-constrained client must verify work performed by an untrusted server without re-executing the computation \cite{zorro}.

Among practical ZKP systems, two dominant paradigms are Zero-Knowledge Succinct Non-interactive ARguments of Knowledge (zk-SNARKs) \cite{znsnark} and Zero-Knowledge Scalable Transparent ARguments of Knowledge (zk-STARKs) \cite{zkstark}. Early zk-SNARKs such as Groth16 \cite{groth16} gained broad adoption because of their succinct proofs, but most practical constructions rely on a trusted setup and on number-theoretic assumptions that are vulnerable to quantum attacks. In contrast, zk-STARKs eliminate trusted setup and rely primarily on collision-resistant hash functions, providing both transparency and post-quantum security \cite{snark01,stark01}. Table~\ref{tab:snark_vs_stark_comparison} summarizes the main differences between the two paradigms.

\begin{table}[h!]
\small
\centering
\caption{Comparison of zk-SNARK and zk-STARK protocols in terms of complexity and security assumptions.}
\label{tab:snark_vs_stark_comparison}
\renewcommand{\arraystretch}{1.2}
\begin{tabularx}{\linewidth}{@{} l >{\centering\arraybackslash}X >{\centering\arraybackslash}X @{}}
\toprule
 & \textbf{zk-SNARK} & \textbf{zk-STARK} \\
\midrule
Prover Complexity & Quasi-Linear & Quasi-Linear \\
Verifier Complexity & Constant & Poly-Logarithmic \\
Proof Size & Succinct & Poly-Logarithmic \\
\midrule
Require Trusted Setup & \checkmark & \xmark \\
Post-quantum Secure & \xmark & \checkmark \\
\bottomrule
\end{tabularx}
\end{table}

The scalability of zk-STARK provers depends heavily on the Fast Reed-Solomon Interactive Oracle Proof of Proximity (FRI), which serves as the underlying polynomial commitment scheme. FRI allows a verifier to check that a committed function is close to a low-degree polynomial with logarithmic communication complexity \cite{stark02}. It operates by recursively folding a large polynomial into smaller instances and checking consistency at each round.

Despite these benefits, FRI remains a major prover bottleneck, often accounting for more than 70\% of total runtime \cite{winterfell}. This overhead is driven by two factors: the high memory bandwidth required to generate and commit large evaluation domains, and the irregular memory access patterns introduced by recursive split-and-fold operations. As a result, general-purpose CPUs and GPUs often struggle to fully utilize their compute resources on FRI workloads, leading to limited performance and energy efficiency \cite{airfri}.

To address this bottleneck, we propose \sys, a novel accelerator architecture implemented on Field-Programmable Gate Arrays (FPGAs). \sys is specifically optimized for the FRI protocol and introduces a custom dataflow architecture that efficiently pipelines the core underlying computations while significantly reducing the memory footprint. Unlike fixed-function accelerators that are rigid in their application, \sys adopts a constraint-driven design methodology. By employing a modular framework, \sys can be adapted to a wide range of FPGA platforms. This flexibility enables extensive design space exploration, allowing our framework to accommodate the varying design parameters required by diverse privacy-preserving applications and to be implemented on different grades of FPGAs. In short, our contributions are as follows:

\begin{itemize}
\item We propose \sys, a novel FPGA accelerator architecture designed for the FRI protocol. To the best of our knowledge, our modular hardware design is the first work that enables flexible configuration across diverse FPGA platforms: from high-performance to resource-constrained devices.

\item We evaluate \sys across multiple FPGA device classes, demonstrating significant performance and efficiency gains. \sys achieves $25.7\times$ and $3.5\times$ speedups over state-of-the-art CPU and GPU implementations, respectively, while delivering superior energy efficiency. 
\item Our open-sourced, parameterizable HLS implementation\footnote{Code will be made available in the final version} features an automated design flow that takes user-provided board specifications to synthesize optimal hardware configurations. By providing a flexible, high-performance kernel that scales across device classes, we lower the barrier for zk-STARK deployment in real-world applications. 

\end{itemize}
\section{Preliminaries and Related Works}

\subsection{Notations}
We establish several important notations and parameters used throughout this paper. We denote the 128-bit and BN254 prime moduli as $p_{128}$ and $p_{254}$, respectively, with their corresponding finite fields represented by $\mathbb{F}_{p_{128}}$ and $\mathbb{F}_{p_{254}}$. To formalize the FRI protocol and its underlying cryptographic primitives, we define $\mathcal{D}$ as the evaluation domain, with size $|\mathcal{D}|$, and $d$ as the polynomial degree, where the FRI blowup factor $\rho$ relates them such that $|\mathcal{D}| = \rho \cdot d$. We denote the Merkle tree depth as $dep$ and the folding factor as $F$. For hardware design constraints, we use $M_{\text{max}}$ for on-chip memory capacity, including both BRAM and URAM, $L_{\text{max}}$ for available logic resources such as LUTs and registers, $DSP_{\text{max}}$ for DSP blocks, and $N_{\text{ch}}$ for the number of available off-chip memory channels.

\subsection{Zero-Knowledge Proofs}
Zero-Knowledge Proofs (ZKPs) enable a prover, \Prv, to convince a verifier, \Vrf, of the validity of a statement without revealing the underlying secret input \cite{zkp,snark01}. Such a statement is typically represented as a computation \Cir satisfying $\mathcal{C}(x,w)=y$, where $x$ and $y$ are public inputs and outputs, and $w$ is the secret witness. The protocol satisfies three fundamental properties:
\begin{itemize}
    \item \textbf{Completeness}: If the statement is true and both \Prv and \Vrf follow the protocol honestly, \Vrf accepts the proof with overwhelming probability.
    \item \textbf{Soundness}: A malicious prover \mPrv cannot convince \Vrf to accept a false statement, except with negligible probability.
    \item \textbf{Zero-Knowledge}: \Vrf learns nothing about the secret witness $w$ beyond the validity of the statement.
\end{itemize}

ZKPs can be categorized into interactive and non-interactive proofs. While early constructions relied on interactive exchanges, modern applications typically require non-interactive proofs to enable public verifiability. Among existing constructions, zk-SNARKs and zk-STARKs are the dominant paradigms for commercial deployment \cite{risc0_2025, zcash}. zk-SNARKs are widely adopted because of their succinct proof sizes and fast verification, but they often rely on elliptic-curve pairings and trusted setup \cite{snarkvsstark}. In contrast, zk-STARKs eliminate trusted setup and rely only on collision-resistant hash functions, providing transparency and post-quantum security \cite{stark02}. As a result, zk-STARKs have been adopted in large-scale applications such as Zero-Knowledge Virtual Machines \cite{zkvm} and ZK-Rollups for high-throughput blockchains \cite{zkrollups}. However, their adoption is hindered by the substantial computational overhead of the FRI protocol, briefly introduced in Sec. \ref{sec:FRI_intro}.
\subsection{Related Work}
Recent works on hardware acceleration for Zero-Knowledge Proofs (ZKPs) has largely followed two directions: accelerating individual cryptographic primitives or designing end-to-end accelerators.

The first direction focuses on optimizing key computational kernels, such as ZKP-friendly hash functions \cite{amaze, hashemall}, multi-scalar multiplications (MSMs) \cite{ifzkp}, and number-theoretic transforms (NTTs) \cite{eminem}. These works deliver important speedups for standalone building blocks, but incorporating such optimized kernels into a complete ZKP pipeline remains nontrivial.

The second direction targets end-to-end ZKP acceleration. Prior efforts have explored Application-Specific Integrated Circuits (ASICs) to achieve state-of-the-art performance \cite{szkp, zkspeed, unizk, pipezk, nocap}. However, ASIC-based designs are inherently inflexible, making them difficult to adapt to rapidly evolving ZKP protocols, while also requiring long development cycles. GPUs have also been used for end-to-end or large-scale proving acceleration \cite{GZKP,cuzk,DistMSM,batchzk}. Although they offer massive parallelism and high throughput in datacenter settings, their high power consumption makes them less suitable for resource-constrained or edge environments.

To bridge this gap, \sys presents a modular architecture for end-to-end FRI acceleration that scales with target hardware constraints. By mapping the full ZKP dataflow onto adaptable and widely accessible reconfigurable hardware, \sys enables practical deployment across a broad range of platforms, from datacenter systems to embedded edge devices.

\section{FRI Overview and Implementation Challenges}\label{sec:FRI_intro}
\begin{figure}
    \centering
    \includegraphics[width=\linewidth]{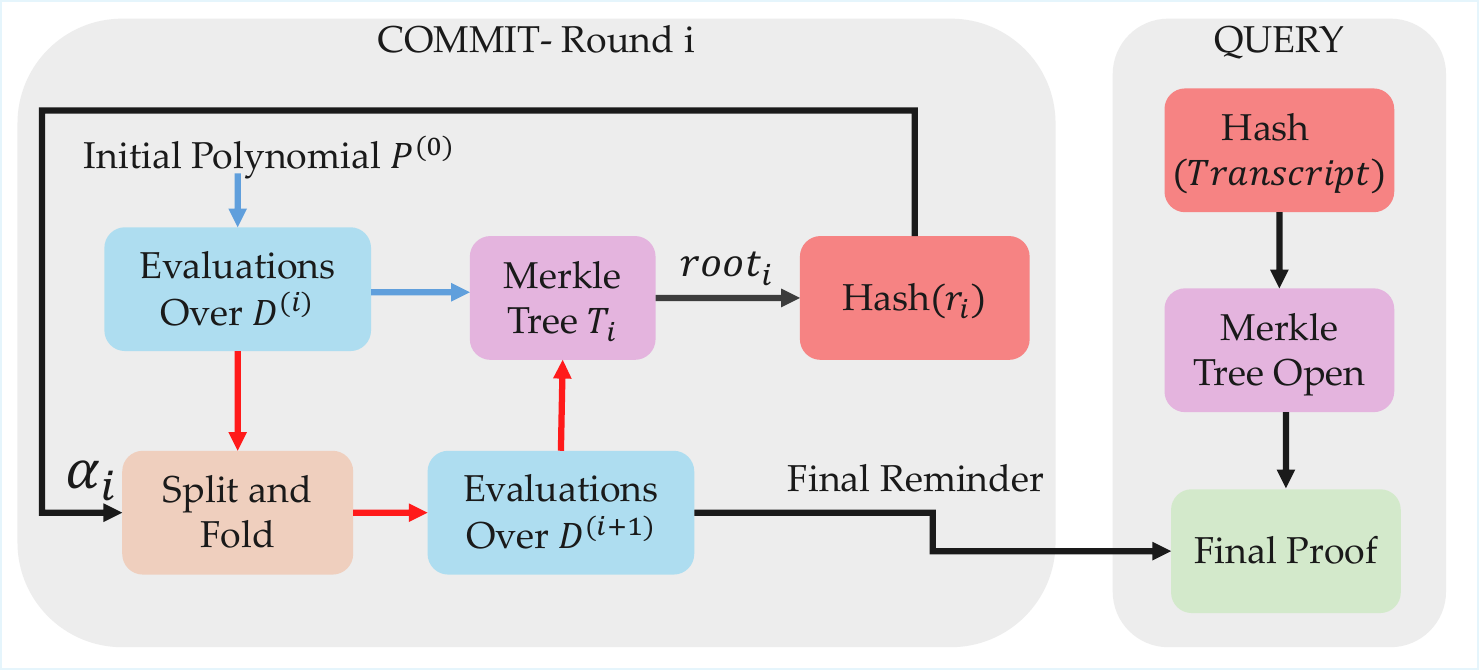}
    \caption{End-to-end FRI workflow. The left panel shows the round-$i$ COMMIT phase: blue arrows denote the initial-round dataflow, red arrows denote subsequent rounds, and black arrows denote common dataflow. The right panel shows the QUERY phase.}
    \label{fig:fri_flow}
\end{figure}

Fast Reed–Solomon Interactive Oracle Proof of Proximity (FRI) protocol is the scalable engine behind zk-STARKs that certifies that a committed function is close to a low-degree polynomial \cite{fri}. It operates over Reed--Solomon codewords on a low-degree extension domain and proceeds in two phases, COMMIT and QUERY, as shown in Fig.~\ref{fig:fri_flow}. In the COMMIT phase, the prover evaluates the base polynomial $P^{(0)}$ over an initial domain $\mathcal{D}^{(0)}$, then enters a recursive loop of vector commitments through Merkle trees and split-and-fold transformations, reducing the problem size until it reaches a small constant. In the QUERY phase, the verifier probabilistically checks folding consistency using Fiat--Shamir-derived random challenges, which the prover traces back through the folding layers to identify the corresponding positions in each intermediate domain $\mathcal{D}^{(i)}$. As the primary bottleneck in zk-STARKs, the FRI protocol presents two major implementation challenges \cite{zksurvy}: high computational overhead from frequent finite-field and hash operations, and memory constraints caused by large data volumes and irregular access patterns.

\subsection{Finite Field Arithmetic}
\label{sec::primefield}

Finite-field arithmetic is a core component of ZKP systems, as all computations are performed modulo a prime $p$, whose choice affects both security and implementation efficiency \cite{zkstark}. In practical systems, $\mathbb{F}_p$ is often defined over a prime of at least 128 bits, so multiplying two field elements $a,b \in \mathbb{F}_p$ yields a double-width result that must be reduced to modulo $p$ \cite{snark01}. Two main approaches are commonly used in practical FRI deployments. The first uses arithmetic-friendly primes, such as $p_{128} = 2^{128} - 45 \cdot 2^{40} + 1$ in Winterfell \cite{winterfell}, whose special form enables efficient reduction with shifts, additions, and small-constant multiplications. The second supports general moduli using methods such as Barrett reduction \cite{amaze,hashemall}. These approaches reflect a tradeoff: structured primes reduce hardware cost but limit flexibility, while general moduli offer broader parameter choices at higher resource cost.

\subsection{Number Theoretic Transformation}
\label{sec:ntt}
Number Theoretic Transformation is the finite-field analog of the Fast Fourier Transform and is widely used to accelerate polynomial evaluation and interpolation over structured domains \cite{ntt}. Given a primitive $n$-th root of unity $\omega_n$, the NTT maps a degree-$(n-1)$ polynomial $P(X)=\sum_{i=0}^{n-1} a_i X^i$ from coefficient form to its evaluations over $\mathcal{D}=\{1,\omega_n,\dots,\omega_n^{n-1}\}$ \cite{ntteval}. In FRI, NTT-based evaluation reduces the cost of constructing Reed--Solomon codewords from $\mathcal{O}(n^2)$ to $\mathcal{O}(n\log n)$ \cite{nttbench}, and the NTT has become a well-studied kernel in ZKP and FHE accelerators \cite{szkp,pipelonk,efficientntt}.

A common technique is the 4-step NTT, which decomposes an $N$-point NTT into smaller $N_1$-point and $N_2$-point NTTs, where $N = N_1 \times N_2$. As shown in Algorithm~\ref{alg:four_step_ntt}, the original vector is reshaped into an $N_1 \times N_2$ matrix. The computation proceeds as column- and row-wise NTTs. This decomposition improves on-chip memory scalability by allowing the design to operate with smaller local buffers and twiddle arrays. Critically, the twiddle-multiplication stage in the middle (line 4-9) introduces three concurrent data dependencies per element: a read of the intermediate matrix entry $a[i][j]$, a read of the twiddle factor $\omega[i \cdot j]$, and a write of the updated product. When these data structures reside in off-chip memory, all three streams must be served simultaneously to sustain throughput, making memory-port allocation a key design concern. In addition, the column-NTT stage accesses data along matrix columns, which becomes a strided access pattern under row-major storage. Efficient implementations must therefore reorganize intermediate data between stages to convert these strided accesses into sequential, burst-friendly patterns.

\begin{algorithm}[htbp!]
\small
    \caption{4-Step NTT Algorithm}
    \label{alg:four_step_ntt}
    \begin{algorithmic}[1]
        \Require 2D array $a[0..n_1-1][0..n_2-1]$, $n = n_1 \times n_2$, twiddle factors $\omega[0..n-1]$
        \Ensure NTT of $a$
        
        \For{$j = 0$ \textbf{to} $n_2 - 1$} \Comment{Column NTTs}
            \State $a[\cdot][j] \gets \mathrm{NTT}_{n_1}(a[\cdot][j])$
        \EndFor
        
        \For{$i = 0$ \textbf{to} $n_1 - 1$} \Comment{Twiddle-factor multiplication}
            \For{$j = 0$ \textbf{to} $n_2 - 1$}
                \State $\omega_{\mathrm{idx}} \gets i \cdot j$
                \State $a[i][j] \gets a[i][j] \cdot \omega[\omega_{\mathrm{idx}}] \bmod q$
            \EndFor
        \EndFor
        
        \For{$i = 0$ \textbf{to} $n_1 - 1$} \Comment{Row NTTs}
            \State $a[i][\cdot] \gets \mathrm{NTT}_{n_2}(a[i][\cdot])$
        \EndFor
        
        \State \Return $a$
    \end{algorithmic}
\end{algorithm}

\subsection{Split-and-Fold Transformation}
\label{subsec:split_fold}
The split-and-fold transformation is the algebraic core of FRI, as it reduces both the polynomial degree and the evaluation-domain size in each round \cite{stark01}. Let $P^{(i)}(X)$ be a polynomial of degree at most $d^{(i)}$ over a field $\mathbb{F}_p$. It can be decomposed into its even and odd components as
\[
P^{(i)}(X) := P^{(i)}_{\mathrm{even}}(X^2) + X \cdot P^{(i)}_{\mathrm{odd}}(X^2),
\]
where $P^{(i)}_{\mathrm{even}}$ and $P^{(i)}_{\mathrm{odd}}$ each have degree at most $d^{(i)}/2$. Given a verifier challenge $\alpha_i \in \mathbb{F}_p$, the fold step combines these two components into a new polynomial
\[
P^{(i+1)}(X) := P^{(i)}_{\mathrm{even}}(X) + \alpha_i \cdot P^{(i)}_{\mathrm{odd}}(X),
\]
which also has degree at most $d^{(i)}/2$. By repeating this process across rounds, FRI progressively transforms the original proximity-testing problem into a much smaller instance.

In practice, it is performed directly on the evaluation vector rather than reconstructing the polynomial explicitly, thereby avoiding expensive interpolation \cite{fri}. Specifically, evaluations at points $x$ and $-x$, which both map to $x^2$, are combined as
\[
P^{(i+1)}(x^2) = \frac{P^{(i)}(x) + P^{(i)}(-x)}{2} + \alpha_i \frac{P^{(i)}(x) - P^{(i)}(-x)}{2x}.
\]
This formulation is especially important for hardware because it exposes the exact arithmetic and data movement required in each folding round. First, the coefficient $(2x)^{-1}$ depends only on the evaluation domain and can therefore be precomputed and stored as an auxiliary inverse-offset table, avoiding repeated inversions during execution. Second, each folding operation consumes a pair of input evaluations, $P^{(i)}(x)$ and $P^{(i)}(-x)$, together with the corresponding precomputed coefficient, and produces one output in the next round. Under a linear memory layout, these paired inputs are logically related but not stored contiguously, which complicates efficient off-chip access. As a result, high-throughput hardware support for split-and-fold depends not only on fast finite-field arithmetic, but also on the memory access that can continuously feed paired evaluations and coefficients without stalling the pipeline.

\subsection{Collision-Resistant Hash Functions and Vector Commitments}
Collision-resistant hash functions are a core primitive in zk-STARKs. They serve two purposes: generating pseudorandom challenges via the Fiat--Shamir transformation and constructing vector commitments over large evaluation vectors \cite{merkle02}. In FRI, these commitments are instantiated as Merkle trees over polynomial evaluation domains. A Merkle tree is a complete binary tree over an input vector of size $N = 2^{dep}$, where each node stores a cryptographic digest \cite{merkle}. The input elements form the leaves, and each internal node is computed recursively as $\text{node}_k = H(\text{node}_{2k} \, \| \, \text{node}_{2k+1})$, until a single root digest is obtained as the commitment to the entire input vector. In FRI, the committed data are the evaluation vectors $P^{(i)}(\mathcal{D}^{(i)})$. Each evaluation is placed at a leaf, and the resulting root digest $T_i$ binds the prover to the full evaluation vector at round $i$. To reveal individual evaluations later, the prover supplies an authentication path \cite{merkle}, namely the ordered sequence of $\log_2 N$ sibling digests needed to recompute and verify the root.

\subsection{Implementation Challenges}
The FRI protocol poses two major implementation challenges: memory scalability and computing parallelism.

{\textbf{Memory Scalability.}}
FRI manipulates large intermediate vectors across NTT, split-and-fold, and Merkle tree construction, quickly exceeding the capacity of on-chip BRAM and URAM. As a result, performance depends not only on computation, but also on how efficiently the design moves and organizes data across memory hierarchies. In NTT, multi-stage accesses and twiddle-factor multiplication stress both local buffering and off-chip bandwidth. In Merkle tree construction, retaining intermediate nodes for later decommitment further increases storage pressure. These issues motivate the batched tiled-transpose 4-step NTT in Sec.~\ref{method::ntt} and the customized memory layouts described in Sec.~\ref{method::split-and-fold} and Sec.~\ref{method::merkle}.

{\textbf{Computing Parallelism.}}
FRI also exposes substantial parallelism in finite-field arithmetic and hash-based vector commitments. Increasing the number of arithmetic and hash units can improve throughput, but it also raises consumption of LUTs, registers, DSPs, and on-chip memory. The central design challenge is therefore not simply to maximize parallelism, but to scale it in a resource-aware manner for different FPGA targets. To address this tradeoff, we develop parameterizable hardware-generation strategies for split-and-fold and Merkle tree construction, in Sec.~\ref{method::split-and-fold} and Sec.~\ref{method::merkle}.
\section{Our Proposed Framework: \sys}
To the best of our knowledge, \sys is the first framework to provide a scalable mapping of end-to-end ZKP workloads across different FPGA tiers. The design methodology of \sys centers on a constraint-driven approach to democratize ZKP acceleration. We provide a high-level architectural abstraction that allows users to generate optimized architectures from target-board resource specifications. By automatically navigating key tradeoffs in computational parallelism, memory hierarchy, and on-chip and off-chip bandwidth, \sys generates high-performance FRI accelerators without requiring expertise in hardware design or ZKP protocols.
\subsection{Fast Polynomial Evaluations with NTT}
\label{method::ntt}
Polynomial evaluation over a large domain is a core operation in FRI and is typically implemented with NTT, as discussed in Sec.~\ref{sec:ntt}. However, even the 4-step NTT in Algorithm~\ref{alg:four_step_ntt} does not guarantee that all intermediate data remain on chip for constrained FPGA platforms. For a $2^{20}$-point NTT with 128-bit field elements, the intermediate matrix alone requires roughly 16\,MB of storage, which can exceed the BRAM/URAM capacity of many devices. As a result, the intermediate matrix and twiddle tables must often be placed in off-chip memory, where naive accesses incur both high latency and poor bandwidth utilization.

To address this limitation, \sys employs the batched tiled-transpose 4-step NTT shown in Algorithm~\ref{alg:batched_tiled_ntt}. We highlighted the key changes relative to the standard 4-step NTT in red: the column-NTT stage is performed on only $B$ columns at a time using the temporary buffer $t$ in lines 1--4, and the buffered results are written back in transposed order in lines 5--8. This modification resolves the main storage challenge of the original 4-step schedule: instead of requiring the full intermediate matrix to be buffered on chip, the design only needs a tile of size $N_1 \times B$. At the same time, the transpose reorganizes the off-chip layout into a row-contiguous form, so that the following twiddle-multiplication and row-NTT stages no longer suffer from column-wise strided accesses.

This tiled-transpose schedule improves both storage efficiency and memory latency. By reducing the on-chip working set to the buffer $t$, the design can fit the computation within the BRAM/URAM budget of the target FPGA. More importantly, by converting data layout into a row-contiguous organization, the twiddle-multiplication and row-NTT stages can use long sequential bursts instead of latency-dominated scattered accesses. The batch size $B$ is therefore chosen as large as possible while ensuring that the buffer remains on chip: a larger $B$ reduces the number of iterations in the batched column-NTT stage, lowering loop overhead and amortizing off-chip access latency, whereas a smaller $B$ allows the same architecture to fit more resource-constrained devices.

\begin{algorithm}[htbp]
\small
    \caption{Batched Tiled-Transpose 4-Step NTT}
    \label{alg:batched_tiled_ntt}
    \begin{algorithmic}[1]
        \Require 2D array $a[0..n_1-1][0..n_2-1]$, buffer $t[0..n_1-1][0..B-1]$, $n = n_1 \times n_2$, twiddle factors $\omega[0..n-1]$
        \Ensure NTT of $a$

        \Statex \textcolor{BrickRed}{\textbf{// Process only $B$ columns at a time to keep the on-chip buffer bounded}}
        \For{$j_{\mathrm{base}} = 0$ \textbf{to} $n_2 - 1$ \textbf{step} $B$}
            \For{$j = 0$ \textbf{to} $B - 1$}
                \State $t[\cdot][j] \gets \mathrm{NTT}_{n_1}(a[\cdot][j_{\mathrm{base}} + j])$
            \EndFor

            \Statex \textcolor{BrickRed}{\textbf{// Write back the buffer to reorganize the off-chip layout}}
            \For{$i = 0$ \textbf{to} $n_1 - 1$}
                \For{$j = 0$ \textbf{to} $B - 1$}
                    \State $a[i][j_{\mathrm{base}} + j] \gets t[i][j]$
                \EndFor
            \EndFor
        \EndFor

        \Statex \textcolor{BrickRed}{\textbf{// After tiling, the following stages consume row-contiguous data}}
        \For{$i = 0$ \textbf{to} $n_1 - 1$}
            \For{$j = 0$ \textbf{to} $n_2 - 1$}
                \State $\omega_{\mathrm{idx}} \gets i \cdot j$
                \State $a[i][j] \gets a[i][j] \cdot \omega[\omega_{\mathrm{idx}}] \bmod q$
            \EndFor
        \EndFor

        \Statex \textcolor{BrickRed}{\textbf{// Row NTTs now avoid the strided column access pattern}}
        \For{$i = 0$ \textbf{to} $n_1 - 1$}
            \State $a[i][\cdot] \gets \mathrm{NTT}_{n_2}(a[i][\cdot])$
        \EndFor

        \State \Return $a$
    \end{algorithmic}
\end{algorithm}

The architecture further adapts to the DDR-bank capability of the target platform by following the access pattern induced by Algorithm~\ref{alg:batched_tiled_ntt}. After the batched column NTTs, the results are written back in transposed order, making the intermediate matrix row-contiguous in off-chip memory for the subsequent twiddle-multiplication and row-NTT stages. In the twiddle-multiplication stage, each row generates three off-chip  streams: input data, twiddle factors, and updated outputs. Ideally, these streams are mapped to independent DDR channels to avoid interference and preserve burst accesses. When fewer DDR banks are available, the same algorithm is retained but the transfers are temporally separated: the design burst-reads the required row data, performs the multiplication from on-chip RAM, and then burst-writes the updated row before the row NTT. In summary, the tiled-transpose schedule remains effective across platforms with different DDR-bank configurations by preserving sequential burst accesses and adapting the same computation flow to available memory bandwidth.

\subsection{Recursive Split and Fold}
\label{method::split-and-fold}
The split-and-fold operation is invoked at every FRI round to recursively reduce the problem size, and it introduces memory-access challenges similar to those of polynomial evaluation. In each iteration, the kernel must read an evaluation pair separated by half of the current domain size, together with the corresponding inverse offset, before computing and writing the folded result. This combination of strided evaluation accesses and sequential coefficient accesses makes memory organization a primary determinant of performance across FPGA platforms.

Our design generation is therefore centered on the memory system. For platforms with sufficient DDR resources, \sys adopts a dual-buffer ping-pong organization for the working set. In each round, the current layer is read from one buffer and the folded results are written to the other, separating memory reads from writes and allowing the fold loop to sustain initial interval (II)=1. Inverse offsets are supplied from a dedicated memory region, and the folded values are also appended to the global evaluation array for later use. As a result, the fully decoupled implementation exposes four useful DDR streams during folding: the source buffer, the inverse-offset stream, the destination buffer, and the global output stream used for the subsequent \texttt{QUERY} phase. This organization avoids read-after-write hazards between adjacent rounds and keeps the arithmetic pipeline decoupled from memory contention.

For more constrained multi-port platforms, \sys uses a more compact layout in which the explicit ping-pong buffers are removed and folded values are written directly into new regions of the evaluation array. Under this organization, the current layer and the next-round layer share the same evaluation channel, while inverse offsets remain on a separate channel. Consequently, the fold arithmetic is reduced to a two-channel layout: one channel serves the evaluation stream, including both source reads and destination writes at disjoint address ranges, and the other serves the inverse-offset stream. Although this organization reduces the overlap available in the fully decoupled design, it preserves pipelined folding and still separates the most critical coefficient traffic from the evaluation traffic.

On platforms constrained to a single DDR interface, the same split-and-fold algorithm is retained, but all evaluation layers and inverse offsets are packed into a unified memory space and distinguished only by disjoint base pointers. Under this organization, the fold loop remains pipelined, but all logical streams are time-multiplexed over the same physical interface, so throughput is bounded by shared DDR bandwidth rather than arithmetic latency. Because the domain size shrinks geometrically across FRI rounds, the corresponding traffic volume also decreases geometrically, which helps keep later rounds efficient even on bandwidth-constrained devices. Overall, DDR-channel count does not change the recursive fold algorithm itself; rather, it determines whether the logical streams of each round can be fully separated, partially compacted, or completely serialized through a shared off-chip path.

\subsection{Merkle Tree Vector Commitment}
\label{method::merkle}
Merkle tree generation constitutes the primary performance bottleneck in the FRI protocol. Unlike polynomial evaluation, which is performed only once per layer, Merkle tree construction is invoked repeatedly across all rounds and therefore dominates total proving time. Consequently, our design prioritizes both the hash compute kernels and their associated memory access patterns.

To address the computational cost, we build specialized high-throughput hash engines by composing primitive round functions from the Vitis Security Library \cite{VitisSecurityLib}. \sys constructs fully pipelined custom hash modules in which the pipeline initiation interval and loop-unrolling factor serve as the primary design knobs. This parameterization allows \sys to adapt automatically to the target device: on high-end FPGAs, hash cores are fully unrolled to execute all rounds in a deep pipeline for maximum throughput, while on resource-constrained platforms, the design uses an iterative architecture to reuse functional units and fit within tight area budgets. By dynamically scaling the hash pipeline, the accelerator saturates the computational capability of a given board without requiring manual RTL redesign.
\begin{figure}[htbp]
    \centering
    \includegraphics[width=\linewidth]{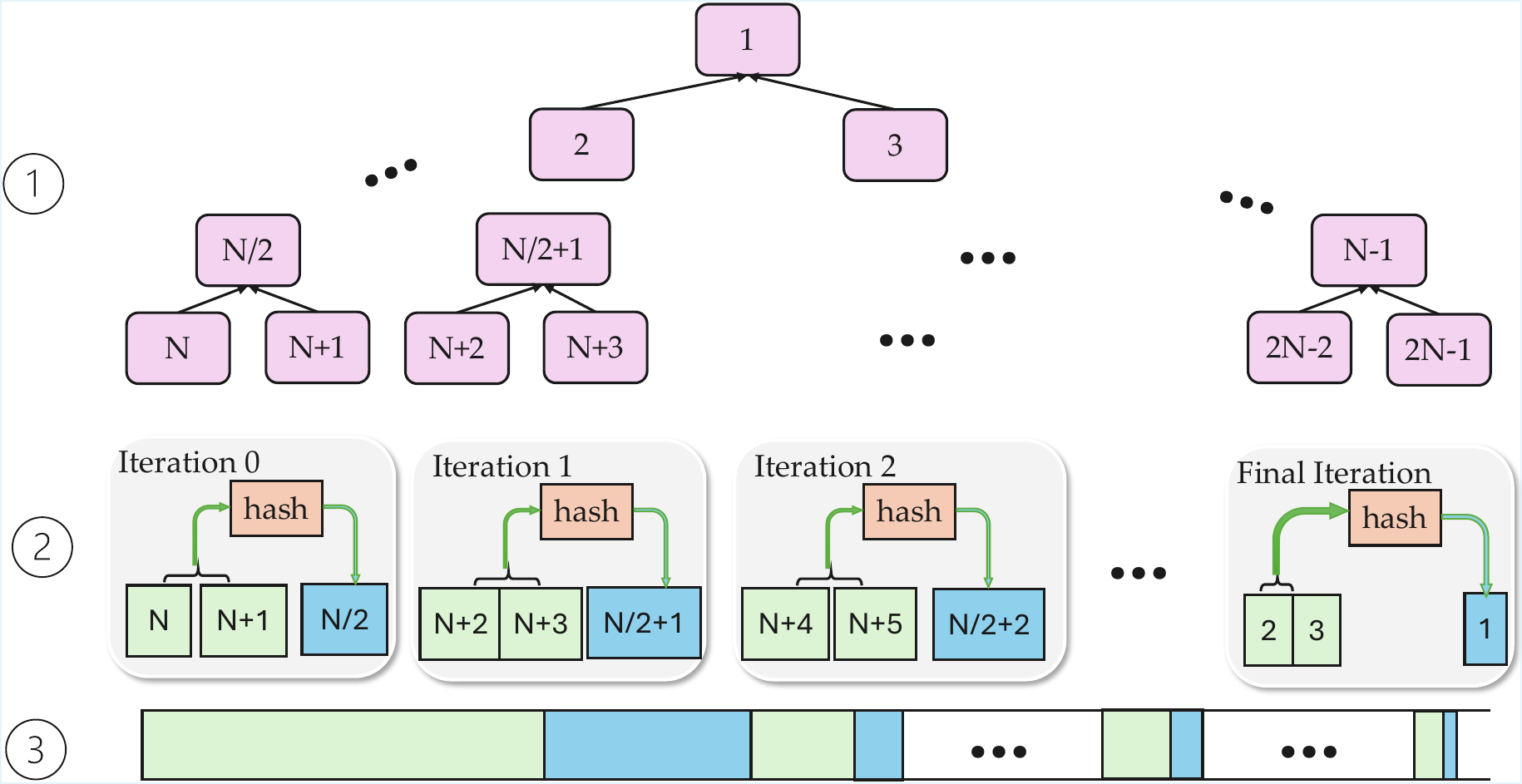}
    \caption{Merkle Tree Memory Architecture. \textcircled{1}: Reversed Linearized mapping of a depth-$d$ tree to flat DDR address space ($N=2^{dep}$). \textcircled{2}: Reverse-order iteration generating contiguous physical addresses for sibling reads and parent writes. \textcircled{3}: Resulting long-stream DDR bursts that maximize off-chip bandwidth utilization.}
    \label{fig:merkle}
\end{figure}
    
Memory management is equally critical and adapts to the storage hierarchy of the target platform. On high-end FPGAs with abundant on-chip RAM, \sys stores the entire Merkle tree internally and employs cyclic array partitioning with a factor of two to create physical banks for even and odd indices, enabling simultaneous sibling fetches and allowing the HLS scheduler to maintain II=1. Conversely, for mainstream or edge devices where all nodes must reside in off-chip DDR, \sys mitigates the high latency of random access by employing a reverse-order linearized binary heap layout. As illustrated in Figure~\ref{fig:merkle}\,\textcircled{1}, the tree is mapped to a 1-based flat array where the leaf nodes occupy the range $[N,2N-1]$. During construction, the engine iterates the parent index $i$ linearly from $N-1$ down to 1, computing $h = \text{Hash}(\text{node}[2i], \text{node}[2i+1])$. As shown in Figure~\ref{fig:merkle}\,\textcircled{2}, because $i$ decrements by one in each iteration, the resulting sibling-read addresses $2N \to 2$ and parent-write addresses $N-1 \to 1$ form strictly contiguous memory streams. This spatial and temporal adjacency, visualized in Figure~\ref{fig:merkle}\,\textcircled{3}, enables the DDR controller to coalesce thousands of individual requests into long, high-efficiency AXI4 bursts. As the tree depth increases, these contiguous streams become longer, allowing the memory controller to saturate available bandwidth and effectively amortize DDR access latency to near zero.

Finally, in the \texttt{QUERY} phase, Merkle tree decommitment traverses individual trees to verify proof authenticity, leading to random accesses from DDR that would otherwise incur high latency penalties. However, decommitment has logarithmic complexity $\mathcal{O}(\log N)$ relative to tree size and requires only a single authentication path per query. The small number of memory accesses makes this latency overhead negligible compared with the \texttt{COMMIT} phase, which dominates total execution time.

\subsection{Design Abstractions and Automatic Hardware Generation}

To efficiently map our FRI architecture to diverse FPGAs, \sys employs an automatic, component-level design space exploration method. It decomposes the architecture into fundamental building blocks: NTT, split-and-fold, and Merkle hash cores, with configurable microarchitectural parameters. Algorithm \ref{alg:fastar_gen} outlines this search process, which aggregates the expected resource footprint for each configuration ($\mathcal{S}_{NTT} \times \mathcal{S}_{SF} \times \mathcal{S}_{Hash}$) and prunes those exceeding the target board's physical limits ($M_{max}$, $L_{max}$, $DSP_{max}$, $N_{ch}$). From the surviving candidates, it selects the configuration yielding the highest computational performance. By automating this search, \sys seamlessly scales across FPGA tiers. We categorize these FPGA tiers as follows: 

\textbf{Datacenter-grade FPGAs} These platforms are characterized by abundant logic and on-chip RAM resources. Given abundant resources, the algorithm prioritizes fully pipelined modules, deep loop unrolling, and independent memory channels for all major traffic classes to maximize throughput and eliminate memory interference.

\textbf{Mainstream High-Performance FPGAs} \sys-M targets mainstream platforms with moderate on-chip memory and fewer memory interfaces. The algorithm adapts by sharing evaluation streams and tuning pipelines to maintain arithmetic saturation via efficient DDR burst accesses.

\textbf{Embedded Edge FPGAs} \sys-E targets edge devices with strict area and bandwidth limits. It maps all large data structures to a unified off-chip layout multiplexed over a single DDR interface, scaling down functional units to a serialized execution model.
\begin{algorithm}[htbp]
\caption{\sys Hardware Search and Generation}
\small
\label{alg:fastar_gen}
\begin{algorithmic}[1]
\Require Target Constraints $\mathcal{P}_{target}(L_{max}, DSP_{max}, M_{max}, N_{ch})$
\Require Component Search Spaces $\mathcal{S}_{NTT}$, $\mathcal{S}_{SF}$, $\mathcal{S}_{Hash}$
\Ensure Optimal Hardware Configuration $\mathcal{A}_{gen}$
\State $\mathcal{A} \gets \emptyset,\; T^* \gets \infty$
\For{each $C \in \mathcal{S}_{NTT} \times \mathcal{S}_{SF} \times \mathcal{S}_{Hash}$}
\State $R_{RAM} \gets c_{ntt}.RAM + R_{fixed}$
\State $R_{DSP} \gets c_{ntt}.DSP + c_{sf}.DSP$
\State $R_{LUT} \gets c_{ntt}.LUT + c_{sf}.LUT + c_{hash}.LUT$
\State $Req_{ch} \gets \max(c_{ntt}.ch, c_{sf}.ch, c_{hash}.ch)$
\If{$R_{RAM} \le M_{max} \land R_{DSP} \le DSP_{max} \land R_{LUT} \le L_{max} \land Req_{ch} \le N_{ch}$}
\State $\textbf{if } T_{FRI}(C) < T^* \textbf{ then } \mathcal{A} \gets C,\; T^* \gets T_{FRI}(C)$
\EndIf
\EndFor
\State $\mathcal{A}_{gen} \gets \mathcal{A}$
\State \Return $\mathcal{A}_{gen}$
\end{algorithmic}
\end{algorithm}
\paragraph{\textbf{Hardware Generation}} To physically realize these designs, our design relies on a custom library of precompiled, HLS templates configured by their dominant resource footprints. NTT templates scale on-chip storage via memory batch sizes. Merkle tree engines wrap optimized Xilinx Vitis Security Library primitives, scaling LUT utilization via pipeline initiation interval (II) and parallel hash core counts. The recursive split-and-fold arithmetic scales DSP usage by tuning its pipeline II. Crucially, all templates are cross-compiled with varying AXI configurations to natively support the specific memory channels required by \sys-D, \sys-M, and \sys-E. Once the search identifies the optimal parameters, the HLS toolchain compiles the selected templates into the final hardware design.

\begin{table*}[!t]
\small
\centering
\caption{Hardware resource utilization, latency, and power of \sys under two prime-field configurations across three FPGA classes: datacenter-grade (\sys-D), mainstream-grade (\sys-M), and embedded-edge (\sys-E)}
\label{tab:resource_perf}
\resizebox{\textwidth}{!}{%
    \setlength{\tabcolsep}{3pt}
    \begin{tabular}{ccccccccccc}
    \toprule
    \multirow{2}{*}{\textbf{Prime Field}} &
    \multirow{2}{*}{\textbf{Implementation}} &
    \multicolumn{5}{c}{\textbf{Resource Utilization}} &
    \multicolumn{2}{c}{\textbf{Timings}} &
    \multirow{2}{*}{\textbf{Power (W)}} \\
    \cmidrule(lr){3-7} \cmidrule(lr){8-9}
    & & \textbf{LUTs} & \textbf{FFs} & \textbf{BRAM} & \textbf{URAM} & \textbf{DSPs}
    & \textbf{Freq. (MHz)} & \textbf{Lat. (s)} & \\
    \midrule
    \multirow{3}{*}{$\mathbb{F}_{p128}$}
        & \sys-D & 678046 (52\%) &  130846 (5\%) & 84 (2\%) & 514 (53\%) &  241 (2\%) & 158 & 0.17& 26.5 \\
        & \sys-M & 327257 (62\%) &  79576 (7\%) & 63 (3\%) & 64 (50\%) &  161 (8\%) & 137 & 0.42 & 10.1 \\
        & \sys-E & 99968 (36\%) & 88671 (16\%) &1056 (57\%) & -- & 161 (6\%) & 91 & 1.36 & 3.4 \\
    \midrule
    \multirow{3}{*}{$\mathbb{F}_{p254}$}
        & \sys-D & 764597 (58\%) & 216658 (8\%) &88 (2\%) & 512 (53\%) & 1338 (14\%) & 143 & 0.19 & 29.6 \\
        & \sys-M & 372960 (71\%) & 94537 (9\%) &  72 (3\%) &64 (50\%) &  676 (34\%)& 125 & 0.51 & 11.2 \\
        & \sys-E & 169468 (61\%) & 173335 (31\%) &1072 (58\%) & -- & 901 (35\%) & 89 & 2.44 & 5.8 \\
    \bottomrule
    \end{tabular}%
}
\end{table*}
\section{Experiments}

\subsection{Setup}
\label{subsec:experimental_setup}

The hardware modules for \sys were developed using a C++ front-end in Xilinx Vitis 2025.2. We evaluated the design across three distinct FPGA tiers. For the \textbf{datacenter class}, we targeted the Virtex UltraScale+ (part number \texttt{xcvu37p-\allowbreak fsvh2892-\allowbreak 2-\allowbreak e}), featuring 1,303,680 LUTs, 2,607,360 FFs, 9,024 DSPs, 2,016 RAM blocks, and 32 HBM channels. For the \textbf{mid-tier FPGA} evaluation, we used the Zynq UltraScale+ MPSoC (part number \texttt{xczu19eg-\allowbreak ffvb1517-\allowbreak 2-\allowbreak e}), equipped with 522,720 LUTs, 1,045,440 FFs, 1,968 DSPs, and 984 RAM blocks. While being significantly smaller than the datacenter device, it retains ample logic resources and four DDR channels. Finally, for the \textbf{embedded-edge tier}, we implemented the design on a Kintex UltraScale+ (part number \texttt{xcku9p-\allowbreak ffve900-\allowbreak 2-\allowbreak i}). With 274,080 LUTs, 548,160 FFs, 2,520 DSPs, and 912 RAM blocks, this device is approximately $2\times$ smaller than the mid-tier FPGA and is constrained by a much more limited off-chip memory interface.

Our primary software baseline is \textbf{Winterfell}, a highly optimized zk-STARK library developed by Meta \cite{winterfell}. We further compare \sys against two GPU baselines, Icicle \cite{icicle} and Air-FRI \cite{airfri}. CPU benchmarks were conducted on a 12th Gen Intel Core i7-12700K processor with 64\,GB of RAM. Icicle was benchmarked on an NVIDIA RTX 3070, while Air-FRI was evaluated on an NVIDIA A40 GPU. All baselines were tested with an FRI domain size of $\mathcal{|D|}=2^{17}$ and a folding factor of $F=2$. The folding process terminated at a final domain size of $2^5$, resulting in a total of 12 rounds. The hash function used by all designs is SHA3. The latency numbers reported in this work reflect complete FPGA-side FRI execution. Host-side trace generation, constraint evaluation, and host--device transfer overheads are excluded.

\subsection{\sys Evaluation}

Table~\ref{tab:resource_perf} demonstrates that the tradeoffs across the three FPGA tiers in term of the degree of memory usage and computing parallelism. \sys-D adopts the most aggressive memory organization and this architecture keeps the main evaluation, folding, and intermediate buffering streams largely decoupled, allowing the latency-critical kernels to sustain an initiation interval (II) of 1 at the highest achieved frequencies among all configurations -- 158\,MHz for $\mathbb{F}_{p128}$ and 143\,MHz for $\mathbb{F}_{p254}$. Consequently, \sys-D achieves the lowest end-to-end latency, albeit at the cost of the largest LUT, URAM, and power footprint. These synthesis results indicate that the datacenter-tier design is optimized primarily for maximum throughput, treating area and energy efficiency as secondary constraints. \sys-M preserves the same pipelined FRI organization but scales the design down by reducing external-memory parallelism and consolidating selected traffic classes onto shared memory bundles. Although this reduces bandwidth isolation compared to \sys-D, the core loops still retain II=1. As a result, \sys-M represents a balanced operating point: it significantly cuts logic, DSP, URAM, and power consumption relative to \sys-D while still delivering strong acceleration. Thus, \sys-M occupies the practical middle ground, targeting platforms where throughput remains critical but must fit within a tighter resource and power budget. \sys-E pushes this scaling trend further by collapsing all off-chip traffic onto a single DDR interface. Unlike the higher-tier designs, which rely on multiple independent memory streams to overlap data movement with computation, \sys-E must serialize a much larger fraction of its memory traffic through one external channel. Consequently, \sys-E becomes memory-bound rather than compute-bound. This results in the lowest operating frequency and the highest latency among the three tiers, but it also yields the smallest power usage. Therefore, \sys-E is best understood not merely as a reduced-performance version of \sys-D, but as a distinct, edge-oriented design that prioritizes low power and deployability over absolute latency.

A second crucial trend highlighted in Table~\ref{tab:resource_perf} is the hardware cost associated with scaling the field size from $\mathbb{F}_{p128}$ to $\mathbb{F}_{p254}$. Across all FPGA tiers, the maximum clock frequency drops only modestly; however, the arithmetic footprint, specifically DSP usage, increases sharply. For instance, DSP utilization jumps from 241 to 1,338 on \sys-D and from 161 to 676 on \sys-M. This aligns with the prime field characteristics discussed in Sec.~\ref{sec::primefield}: $\mathbb{F}_{p128}$ is "reduction-friendly," as the operation is primarily performed via bitshifts and small multiplications. On FPGAs, these are typically synthesized using logic cells like LUTs rather than DSPs. Conversely, the $\mathbb{F}_{p254}$ field requires Barrett reduction with 256-bit multiplications, leading to high DSP consumption. This impact is most pronounced in the edge configuration; the latency of \sys-E nearly doubles for $\mathbb{F}_{p254}$. This demonstrates how wider-field arithmetic compounds the bandwidth bottlenecks inherent in a single-channel DDR architecture. In contrast, the Virtex-based \sys-D maintains a nearly constant cycle count of across both fields, proving that a better memory subsystem and compute capacity can effectively absorb the arithmetic overhead.

Overall, these results validate the constraint-driven generation strategy of \sys. While all three implementations share the same tiled NTT structure and modular FRI pipeline, they differ significantly in how memory channels, buffering resources, and pipeline overlaps are provisioned to match target FPGA capabilities. Performance scales predictably with available memory bandwidth and arithmetic resources: \sys-D maximizes throughput via memory- and compute-level parallelism, \sys-M trades a portion of that parallelism for a smaller resource footprint, and \sys-E achieves low power for edge deployments.

\begin{table}[htbp]
\small
\centering
\caption{Comparison of runtime and speedup of \sys on different FPGA grades against a highly optimized CPU baseline. The evaluation is performed in $\mathbb{F}_{p128}$. The CPU and FPGA platforms is described in Sec.~\ref{subsec:experimental_setup}.}
\label{tab:cpu_comparison}
\setlength{\tabcolsep}{9pt} 
\begin{tabular}{cccc}
\toprule
\textbf{Design} & \textbf{Platform} & \textbf{Lat. (s)} & \textbf{Speedup} \\
\midrule
Winterfell~\cite{winterfell} & Intel i7 & 4.37 & -- \\
\midrule
\sys-D & Virtex & 0.17  & 25.7$\times$ \\
\sys-M & Zynq   & 0.42  & 10.4$\times$ \\
\sys-E & Kintex & 1.36  & 3.2$\times$ \\
\bottomrule
\end{tabular}
\end{table}
We compare the runtime and speedup of \sys and Winterfell library in Table~\ref{tab:cpu_comparison}. The data shows that all three FPGA tiers deliver meaningful acceleration at entirely different operating points. \sys-D provides the highest performance, reducing the CPU's 4.37\,s latency to just 0.17\,s, yielding a 25.7$\times$ speedup. \sys-M achieves a 10.4$\times$ speedup, confirming that substantial acceleration is retained even when scaling down to a resource-constrained mainstream FPGA. Importantly, the highly bandwidth-limited \sys-E still reduces runtime to 1.36\,s with a 3.2$\times$ speedup, demonstrating that FPGA acceleration remains viable and beneficial even at the embedded edge where power constraints rule out larger devices. 

Taken together, the CPU comparisons highlight the scalable nature of \sys. Performance degrades gracefully from \sys-D to \sys-E as resources are restricted, directly mirroring the synthesis results: as memory parallelism and buffering shrink, latency increases, yet the hardware maintains enough parallel execution to comfortably outperform optimized software.

\begin{table}[htbp]
\small
\centering
\caption{Comparison of the runtime and speedup of \sys across different FPGA grades against GPU implementations over $\mathbb{F}_{p254}$. Both GPU platforms are NVIDIA devices, and the FPGA platforms are described in Sec.~\ref{subsec:experimental_setup}.}
\label{tab:comparison_merged}
\resizebox{\columnwidth}{!}{%
\begin{tabular}{ccccc}
\toprule
\multirow{2}{*}{\textbf{Design}} &
\multirow{2}{*}{\textbf{Platform}} &
\multirow{2}{*}{\textbf{Lat. (s)}} &
\multicolumn{2}{c}{\textbf{Speedup}} \\
\cmidrule(lr){4-5}
& & & \textbf{vs. Icicle} & \textbf{vs. Air-FRI} \\
\midrule
Icicle \cite{icicle}  & RTX 3070   & 0.33  & -- & -- \\
Air-FRI \cite{airfri} & A40 Ampere & 0.67  & -- & -- \\
\midrule
\sys-D & Virtex & 0.19  & 1.69$\times$ & 3.5$\times$ \\
\sys-M & Zynq   & 0.51  & 0.64$\times$ & 1.31$\times$ \\
\sys-E & Kintex & 2.44  & 0.13$\times$ & 0.27$\times$ \\
\bottomrule
\end{tabular}%
}
\end{table}

Table~\ref{tab:comparison_merged} compares \sys with GPU baselines over $\mathbb{F}_{p254}$. Enabled by its specialized architectural optimizations, \sys-D clearly outperforms the GPU implementations, achieving 1.69$\times$ and 3.5$\times$ speedups over Icicle and Air-FRI, respectively. Although \sys-M and \sys-E deliver lower performance due to their more constrained compute resources and memory bandwidth, they occupy an important design point for edge deployment, where integrating discrete GPUs may be impractical because of power or cost constraints.

In summary, the CPU and GPU comparisons illustrate the complete performance envelope of \sys. Against CPUs, all tiers provide substantial speedups, proving the architecture's effectiveness even on constrained devices. Against GPUs, the datacenter design proves FPGAs can beat state-of-the-art baselines, while the mainstream and edge tiers broaden the deployment landscape by offering viable alternatives where GPUs are less efficient or entirely infeasible. These results reinforce our goal: a single hardware framework can scale seamlessly across datacenter, mainstream, and edge FPGAs to expose distinct, optimized points in the latency, power, and resource design space.
\section{Conclusions}
This paper proposes \sys, a novel FPGA-based accelerator for the FRI protocol in post-quantum zk-STARK applications. \sys features an automated design flow that generates optimal hardware configurations from user-provided board specifications, enabling agile deployment across diverse FPGA platforms without manual hardware redesign. Our evaluation demonstrates up to $25.7\times$ and $3.5\times$ speedups over CPU and GPU baselines with superior energy efficiency. In conclusion, \sys features automated hardware generation and an open-source modular design and achieved state-of-the-art performance, making the deployment of post-quantum zk-STARK practical and scalable for real-world applications.
\section*{Acknowledgements}
This work is supported in part by the U.S. National Science Foundation under Grants CNS-2153690, CNS-2247892, CNS-2239672, OAC-2319962, and CNS-2326597. 
\bibliographystyle{ACM-Reference-Format}
\bibliography{sample-base}
\end{document}